# Single-material MoS$_2$ thermoelectric junction enabled by substrate engineering

Mohammadali Razeghi[1], Jean Spiece[3], Oğuzhan Oğuz[1], Doruk Pehlivanoğlu[2], Yubin Huang[3], Ali Sheraz[2], Phillip S. Dobson[4], Jonathan M. R. Weaver[4], Pascal Gehring[3], T. Serkan Kasırga[1,2*]

[1] Bilkent University UNAM – Institute of Materials Science and Nanotehcnology, Bilkent 06800 Ankara, Turkey

[2] Department of Physics, Bilkent University, Bilkent 06800 Ankara, Turkey

[3] IMCN/NAPS, Université Catholique de Louvain (UCLouvain), 1348 Louvain-la-Neuve, Belgium

[4] James Watt School of Engineering, University of Glasgow, Glasgow G12 8LT, U.K

*Corresponding Author: kasirga@unam.bilkent.edu.tr

**Abstract**

To realize a thermoelectric power generator, typically a junction between two materials with different Seebeck coefficient needs to be fabricated. Such difference in Seebeck coefficients can be induced by doping, which renders difficult when working with two-dimensional (2d) materials. Here, we employ substrate effects to form a thermoelectric junction in ultra-thin few-layer MoS$_2$ films. We investigated the junctions with a combination of scanning photocurrent microscopy and scanning thermal microscopy. This allows us to reveal that thermoelectric junctions form across the substrate-engineered parts. We attribute this to a gating effect induced by interfacial charges in combination with alterations in the electron-phonon scattering mechanisms. This work demonstrates that substrate engineering is a promising strategy to develop future compact thin-film thermoelectric power generators.

**Main Text**

In ultra-thin materials with large surface-to-bulk ratio, interactions with the substrate can have strong impact on the materials properties [1–6]. It is therefore important to understand this so-called substrate-effect, especially in order to optimize the reliability of future devices based on two-dimensional (2d) semiconducting materials. As an example, the choice of substrate for mono- and few-layer MoS$_2$ has been shown to strongly affect its Raman modes and photoluminescence (PL)[7], electronic[8], and thermal transport[9] properties. In this work, we employ the substrate effect to enable completely new functionalities in a 2d semiconductor device. To this end, we engineer the substrate that atomically thin MoS$_2$ is deposited on. Using a combination of scanning photocurrent microscopy (SPCM) along with scanning thermal microscopy (SThM) we demonstrate that substrate engineering is a powerful way to build a thermoelectric junction.

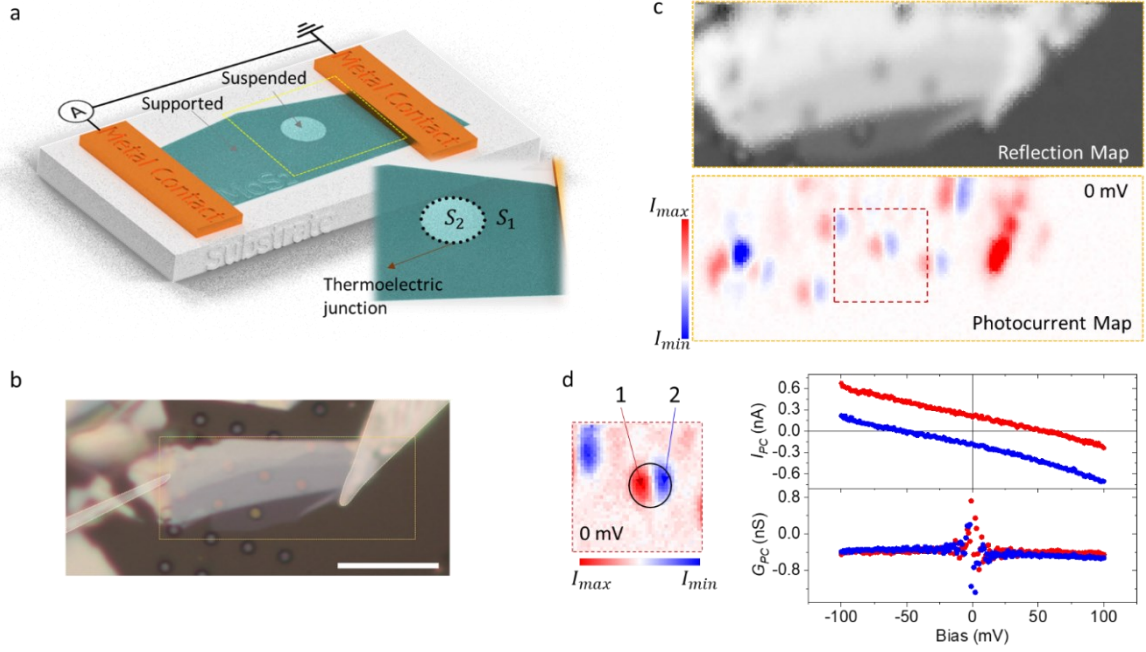

**Figure 1 a.** Schematic of a substrate-engineered device: a MoS$_2$ flake is suspended over a circular hole drilled in the substrate. Metal contacts are used for scanning photocurrent microscopy (SPCM), scanning thermal gate microscopy (SThGM) and *I-V* measurements. The inset shows a magnification of the area indicated by the dashed yellow square, where Seebeck coefficients of supported and suspended parts are labelled with $S_1$ and $S_2$, respectively. **b.** Optical microscope image of a multi-layered device over circular holes with indium contacts, marked with grey overlays. Scale bar: 10 µm. **c.** SPCM reflection map and the corresponding open-circuit photocurrent map acquired from the yellow dashed rectangle in **b** with 532 nm laser. $\{I_{min}, I_{max}\} = \{-0.5, 0.5\}$ nA. **d.** Photocurrent map from the red dashed rectangle region in **c**. Black circle is the position of the hole determined from the reflection image. Right panel shows the photocurrent, $I_{PC}$ vs bias taken from point 1 (red dots) and point 2 (blue dots) over the suspended part of the crystal marked on the left panel. Lower graph is the derived photoconductance, $G_{PC}$ vs. bias.

In the following we predict that a thermoelectric junction with a Seebeck coefficient difference of tens of µV/K can be fabricated when connecting regions of suspended MoS$_2$ to supported regions. We assume that the Seebeck coefficient $S$ in thermal equilibrium is composed of contributions from the energy-dependent diffusion ($S_N$), scattering ($S_\tau$) and the phonon-drag ($S_{pd}$), so that $S = S_N + S_\tau + S_{pd}$ [9,10]. Here, $S_N$ and $S_\tau$ terms can be written from the Mott relation assuming that MoS$_2$ is in the highly conductive state and electrons are the majority carriers:

$$S_\tau = -\frac{\pi^2 k_B^2 T}{3e} \frac{\partial \ln \tau}{\partial E}\bigg|_{E=E_F} \text{ and } S_N = \pm \frac{k_B}{e}\left[\frac{E_F - E_C}{k_B T} - \frac{(r+2)F_{r+1}(\eta)}{(r+1)F_r(\eta)}\right]$$

where $T$ is the temperature, $k_B$ is the Boltzmann constant, $e$ is the electron's charge, $\tau$ is the relaxation time, $E_F$ is the Fermi energy, $E_C$ is the conduction band edge energy, $r$ is scattering parameter and $E$ is the energy. $F_m(\eta)$ is the *m*-th order Fermi integral[11]. In the 2d limit, $\tau$ is energy independent, thus $S_\tau$ is zero. $S_{pd}$ term can be estimated from the theory of phonon-drag in semiconductors in the first order as $S_{pd} = -\frac{\beta v_p l_p}{\mu T}$ where, $v_p$ and $l_p$ are the group velocity and the mean free path of a phonon, $\beta$ is a parameter to modify the electron-phonon interaction strength and ranges from 0 to 1, and $\mu$ is the electron mobility, respectively[10]. Importantly, $l_p$ and $\mu$ are heavily affected by the presence of a substrate[12] which implies that the $S_{pd}$ term gets strongly modified when the MoS$_2$ flake is suspended.

Indeed, we find that for suspended MoS$_2$ at room temperature $S_{pd} \approx -100$ µV/K and for MoS$_2$ on SiO$_2$ at room temperature $S_{pd} \approx -230$ µV/K. Similarly, $S_N$ is heavily influenced by the presence or absence of the substrate as electron density depends on the interfacial Coulomb impurities and short-ranged defects[11–17]. We estimate that that for MoS$_2$, $S_N$ ranges from -400 µV/K to -200 µV/K for carrier concentrations ranging from 10$^{12}$ cm$^{-2}$ (suspended few layer MoS$_2$) to 3 x 10$^{13}$ cm$^{-2}$ (SiO$_2$ supported few layer MoS$_2$).[18–20] As a result, a substrate engineered thermoelectric junction with a Seebeck coefficient difference of $\Delta S \approx 70$ µV/K can be formed along the MoS$_2$ flake (see **Figure 1a** and Supporting Information).

To test this hypothesis, we fabricated substrate-engineered MoS$_2$ devices by mechanical exfoliation and dry transfer[21] of atomically thin MoS$_2$ flakes on substrates (sapphire or oxidized silicon) with pre-patterned trenches/holes formed by focused ion beam (FIB). We contacted the flakes with Indium needles[22–24] which are suitable for achieving Ohmic contacts to MoS$_2$[25,26] (gold-contacted device measurements are shown in Supporting Information). A typical device is shown in **Figure 1b**. We then used scanning photocurrent microscopy, to locally heat up the junction with a focused laser beam and to measure the photothermoelectric current that is generated (see Methods for experimental details). **Figure 1c** shows the greyscale reflection intensity map and the corresponding photocurrent distribution over the device. For the few-layer suspended MoS$_2$ devices we observe a bipolar photoresponse at the junctions between the supported and the suspended part of the crystal. The spatial distribution of the signal agrees well with the finite element analysis simulations, given in the supporting information, and suggests the formation of a thermoelectric junction. When applying a voltage bias $V$ to the junction, the photocurrent, $I_{PC}$ changes linearly with bias, while the photoconductance, $G_{PC} = \frac{I_{PC}^V - I_{PC}^0}{V}$ ($I_{PC}^V$, $I_{PC}^0$: photocurrent under $V$ and 0 mV bias, respectively) stays constant (**Figure 1d**). Such bias-independent photoconductance is typically an indication for an photothermoelectric nature of the observed signal[22,24,27–29]. Although we propose that the photocurrent in substrate-engineered MoS$_2$ devices is dominated by the photothermal effect (PTE)[30,31], other possible mechanisms have been reported that may lead to a photovoltaic response. These include (1) strain related effects such as strain modulation of materials properties and flexo-photovoltaic effect[13], and (2) substrate proximity related effects that forms a built-in electric field[32].

Next, we present experimental evidence for a thermoelectric origin of the observed photocurrent. To this end, we employed scanning thermal gate microscopy (SThGM), where a hot AFM tip heats up the junction locally while the resulting voltage build-up on the devices is recorded (see Methods). Since no laser-illumination of the sample is required in this method, it can be used to ultimately exclude photovoltaic effects. **Figure 2** compares SPCM and SThGM maps of the same holes. We observed the same bipolar signals in the suspended regions with both experimental methods. Thanks to its sub-100 nm lateral resolution, SThGM further allows us to observe local variations of the thermovoltage in supported MoS$_2$ that can be attributed to charge puddles induced by local doping via the substrate[33–35]. We confirmed that the SThGM signal disappears when no power is dissipated in the probe heater, which rules out parasitic effects induced by the laser used for AFM feedback. Furthermore, SThGM allows us to estimate the magnitude of the local Seebeck coefficient variations. Using the probe-calibration data we obtain a value of $\Delta S = 72 \pm 10$ µV/K (See supporting information). Despite the uncertainties regarding the real sample temperature, the obtained $\Delta S$ value is very close to the theoretically predicted value.

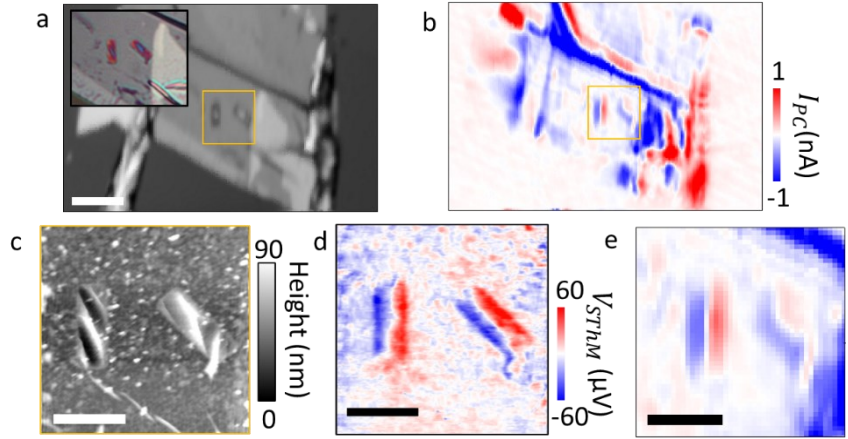

**Figure 2 a.** SPCM reflection map and **b**. photocurrent map of the device shown in the inset of panel ***a***. Scale bar: 10 µm. The yellow rectangle indicates the region that was investigated by SThGM in ***c*** *(AFM height map)* and ***d*** *(SThGM thermovoltage map)*. **e.** SPCM map of the same region excerpted from the map given in ***b***. Color scale is the same as in panel ***b***. Scale bars in ***c***, ***d*** and ***e***: 3 µm.

To understand why suspending MoS$_2$ alters its Seebeck coefficient, we first would like to discuss the possibility of strain induced changes in the materials properties. MoS$_2$, like graphene, is nominally compressed when deposited on a substrate[36–39]. Upon suspending the crystals, the free-standing part either adheres to the sidewalls of the hole and dimples or, bulges. As a result, strain might be present in the free-standing part of the crystal. Strain can affect both the bandgap and the Seebeck coefficient of MoS$_2$. The indirect optical gap is modulated by -110 meV/%-strain for a trilayer MoS$_2$[36,40]. *Ab initio* studies show a ~10% decrease in the Seebeck coefficient of monolayer MoS$_2$ per 1% tensile strain [41]. To estimate the biaxial strain, we performed atomic force microscopy (AFM) height trace mapping on the samples. Most samples, regardless of the geometry of the hole exhibit slight bulging of a few nanometers. For the MoS$_2$ flakes suspended on the circular holes in the device shown in **Figure 3a**, the bulge height is $\delta t \approx 25$ nm. Similar $\delta t$ values were measured for other devices. The biaxial strain can then be calculated using an uniformly loaded circular membrane model, and is as low as 0.0025% [42]. Such a small strain on MoS$_2$ is not sufficient to induce a significant change in bandgap or Seebeck coefficient [43–45].

Next, we consider the substrate induced changes on the material properties. The presence or the absence of the substrate can cause enhanced or diminished optical absorption due to the screening effects, Fermi level pinning[46] and charges donated by the substrate[7,47]. More significantly, the doping effect due to the trapped charges at the interface with the substrate can locally gate the MoS$_2$ and modify the number of charge carriers[48] and thus its Seebeck coefficient. To investigate the electrostatic impact of the substrate on the MoS$_2$ membrane, we investigated the surface potential difference (SPD) on devices using Kelvin Probe Force Microscopy (KPFM). SPD can provide an insight on the band bending of the MoS$_2$ due to the substrate effects[49]. **Figure 3b-d** shows the AFM height trace map and the uncalibrated SPD map of the sample. SPD across the supported and suspended part of the flake is on the order of 50 mV. This shift in the SPD value hints that there is a slight change in the Fermi level of the suspended part with respect to the supported part of the crystal. The same type of charge carriers is dominant on both sides of the junction formed by the suspended and supported parts of the crystal. The band structure formed by such a junction in zero bias cannot be used in separation of photoinduced carriers[50], however, it can lead to the formation of a thermoelectric junction[11,51]. This is in line with the SThGM measurements.

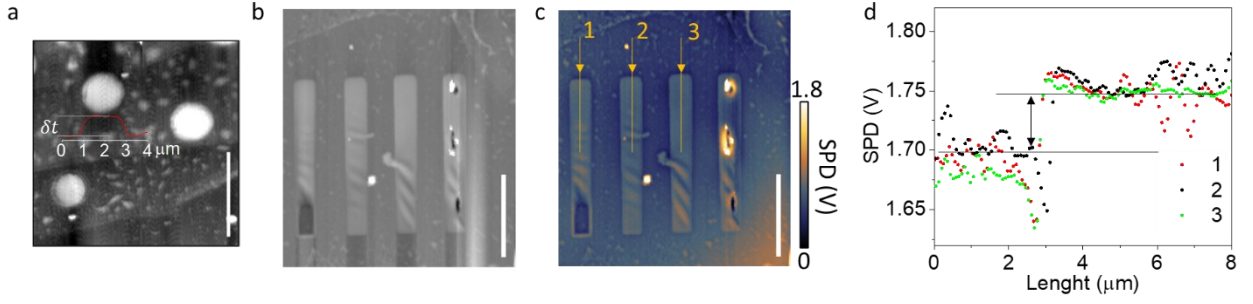

**Figure 3 a.** AFM height trace map of a device suspended over circular holes show a bulge of $\delta t \approx 25$ nm. The line trace is overlayed on the map. Scale bar: 4 µm. **b.** AFM height trace map of the sample shows the bulged and dimpled parts of the flake. Scale bar: 4 µm. **c.** KPFM map of the sample shows the variation in the surface potential. Scale bar: 4 µm. **d.** Line traces taken along the numbered lines in **c**. Direction of the arrows in **c** indicates the direction of the line plot.

In the remainder of the paper, we aim at controlling the electrostatics that are responsible for the formation of a thermoelectric junction. Charge transport in MoS$_2$ is dominated by electrons due to unintentional doping[52,53]. Modulating the density and the type of free charge carriers can be done by applying a gate voltage $V_g$ to the junction[54]. This significantly modifies the magnitude and the sign of the Seebeck coefficient as demonstrated in previous studies[16,30,31,55]. The Mott relation[56] can be used to model the Seebeck coefficient as a function of $V_g$:

$$S = \frac{\pi^2 k_B^2 T}{3e} \frac{1}{R} \frac{dR}{dV_g} \frac{dV_g}{dE}\bigg|_{E=E_F} \quad \text{eq.(1)}$$

Here, $T$ is the temperature, $k_B$ is the Boltzmann constant, $e$ is the electron's charge, $R$ is the device resistance, $E_F$ is the Fermi energy and $E$ is the energy.

Since hole transport is limited due to substrate induced Fermi level pinning on SiO$_2$ supported MoS$_2$ field-effect devices,[46] to observe the sign inversion of the Seebeck coefficient (see the Supporting Information for measurements on device fabricated on SiO$_2$ and Al$_2$O$_3$ coated SiO$_2$) we followed an alternative approach to emulate suspension: we fabricated heterostructure devices where the crystal is partially supported by hexagonal boron nitride (h-BN). h-BN is commonly used to encapsulate two-dimensional materials thanks to its hydrophobic and atomically smooth surface. This leads to less unintentional doping due to the interfacial charge trapping and reduced electron scattering[7,57,58]. A ~10 ML MoS$_2$ is placed over a 10 nm thick h-BN crystal to form a double-junction device (see supporting information for a single-junction device formed by a MoS$_2$ flake which is partially placed over a h-BN flake) and indium contacts are placed over the MoS$_2$. The device is on 1 µm thick oxide coated Si substrate where Si is used as the back-gate electrode. **Figure 4a** shows the optical micrograph of the device and its schematic. The presence of h-BN modifies the SPD by 80 mV – a value very similar to the values we find for suspended devices (see SI) – which is consistent with the relative n-doping by the h-BN substrate[32,57]. We therefore attribute this difference to the Fermi level shift due to the difference in interfacial charge doping by the different substrates.

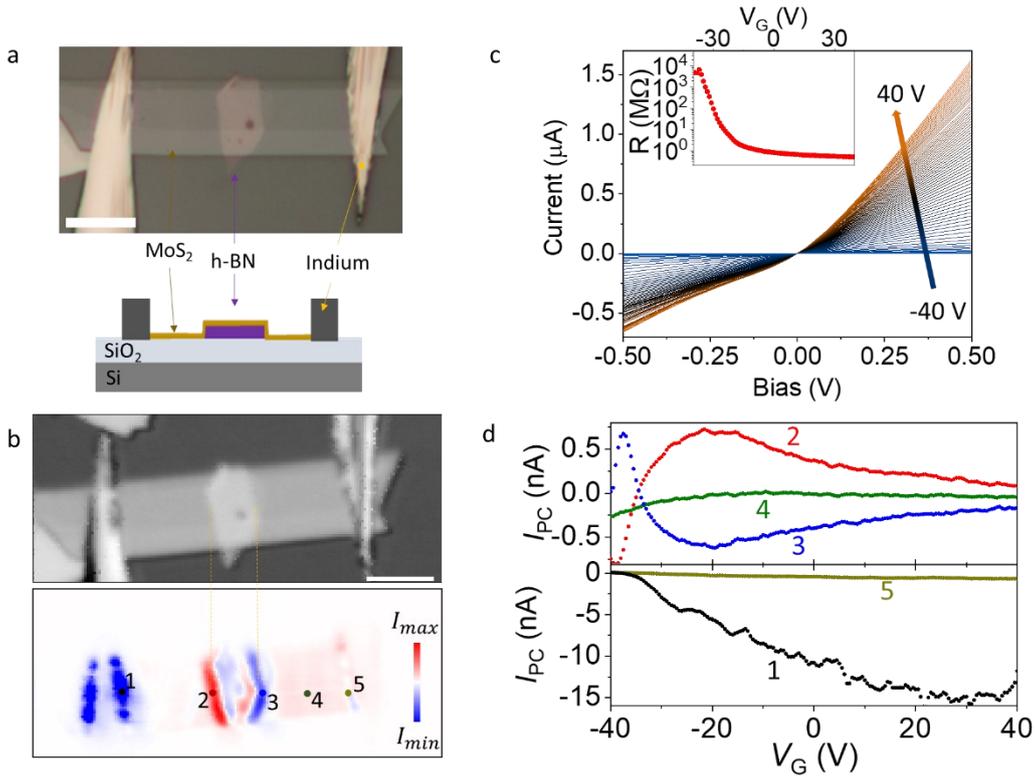

**Figure 4 a.** Optical micrograph of a Si back-gated MoS$_2$ device partially placed over h-BN. Its cross-sectional schematic is shown in the lower panel. Scale bar: 10 µm. **b.** SPCM reflection map and the photocurrent map of the device shown in **a**. $I_{max} = 3$ nA and $I_{min} = -3$ nA. Scale bar: 10 µm. **c.** Current-Voltage graph versus $V_G$ from -40 to 40 V. Inset shows the resistance versus $V_G$. **d.** $I_{PC}$ vs. $V_G$ recorded at the points marked in the SPCM map in **b**.

**Figure 4b** shows the SPCM map under zero gate voltage. We observe a bipolar photocurrent signal from the junctions between h-BN and SiO$_2$ supported MoS$_2$. Raman mapping (see the Supporting Information) reveals slight intensity decrease and a small shift of the $A_{1g}$ peak over the h-BN supported part of the MoS$_2$. This is consistent with the stiffening of the Raman mode due to the higher degree of charged impurities in SiO$_2$ as compared to h-BN[7]. By applying a gate voltage to the device, its resistance can be tuned significantly as free charges are depleted (**Figure 4c**). Under large positive gate voltages, the *I-V* characteristic becomes asymmetric. To investigate the dependence of the photocurrent on carrier type and concentration, the laser is held at specific positions on the device as marked in **Figure 4d**, and the gate is swept from positive to negative voltages with respect to the ground terminal. For positive gate voltages, the magnitude of the photoresponse from both junctions, between h-BN and SiO$_2$ supported MoS$_2$, (points 2 and 3) decrease. When a negative gate voltage is applied, the magnitude of the photoresponse at both junctions increases by almost a factor of two at $V_G = -21.5$ V. Once this maximum is reached, the amplitude of the photocurrent at both points decreases and has the same value as the photocurrent generated over the MoS$_2$ (point 4) at $V_G = -34.5$ V.

These observations can be qualitatively explained as follows: at a gate voltage of $V_G = -34.5\ V$, the majority charge carrier type in the h-BN supported part changes from electrons to holes. As a consequence, the Seebeck coefficients of MoS$_2$ resting on h-BN and SiO$_2$, respectively, become similar, which leads to $\Delta S \approx 0$, and curves 2,3 and 4 in Figure 4d cross. The photocurrent signal recorded near the indium contacts (points 1 and 5) decreases non-monotonically with decreasing $V_G$ and reaches zero at $V_G = -40\ V$. At this voltage the Seebeck coefficient of MoS$_2$ on SiO$_2$ reaches that of Indium ($S_{In}$ = + 1.7 μV/K)[59].

In conclusion we demonstrated that substrate engineering can be used to generate a thermoelectric junction in atomically thin MoS$_2$ devices. Similar strategies can be employed in other low dimensional materials that exhibit large and tunable Seebeck coefficients. This might in particular be promising at low temperature where effects like band-hybridization and Kondo scattering can produce a very strong photothermoelectric effect[9].

**Author Contributions**

T.S.K. designed and conceived the experiments, T.S.K. and P.G. prepared the manuscript. M.R. fabricated devices, performed the experiment and analyzed the results. D.P. prepared the substrates, performed simulations, and helped with the experiments. O.O. performed the AFM and KPFM measurements and A.S. performed some of the earlier measurements. J.S., Y.H. and P.G. performed the SThGM measurements and analyzed the results. P.S.D and J.M.R.W contributed discussions on the implementation of VITA-DM-GLA-1 SThM probes. All authors discussed the results and reviewed the final version of the manuscript.

**Competing Interests**

The Authors declare no Competing Financial or Non-Financial Interests.

**Methods**

SPCM setup is a commercially available setup from LST Scientific Instruments Ltd. which offers a compact scanning head with easily interchangeable lasers. Two SR-830 Lock-in amplifiers are employed, one for the reflection map and the other for the photocurrent/voltage measurements. In the main text we reported the photocurrent (a measurement of the photovoltage is given in **Figure S2**). The incident laser beam is chopped at a certain frequency and focused onto the sample through a 40x objective. The electrical response is collected through gold probes pressed on the electrical contacts of the devices and the signal is amplified by a lock-in amplifier set to the chopping frequency of the laser beam. Although various wavelengths (406, 532, 633 nm) are employed for the measurements, unless otherwise stated we used 532 nm in the experiments reported in the main text (see **Figure S3** for SPCM measurements with different wavelengths). All the excitation energies are above the indirect bandgap of the few layer MoS$_2$.

Scanning Thermal Microscopy measurements were performed with a Dimension Icon (Bruker) AFM under ambient conditions. The probe used in the experiments is VITA-DM-GLA-1 made of a palladium heater on a silicon nitride cantilever and tip. The radius is typically in the order of 25-40 nm. The heater is part of a modified Wheatstone bridge and is driven by a combined 91 kHz AC and DC bias, as reported elsewhere. The signal is detected via a SR830 lock-in amplifier and fed in the AFM controller. This signal monitors the probe temperature and thus allows to locally map the thermal conductance of the sample. In this work, the power supplied to the probe gives rise to a 45K excess temperature.

While the probe is scanning the sample, we measure the voltage drop across the device using a low noise preamplifier (SR 560). This voltage is created by the local heating induced by the hot SThM tip. It is then fed also to the AFM controller and recorded simultaneously. In this study, the thermovoltage measurements were performed without modulating the heater power. We note that it is also possible to generate similar maps by varying the heater temperature and detecting thermovoltage via lock-in detection.

**Data Availability**

Source data available from the corresponding authors upon request.

# Supporting Information: Single-material MoS₂ thermoelectric junction enabled by substrate engineering

Mohammadali Razeghi[1], Jean Spiece[3], Oğuzhan Oğuz[1], Doruk Pehlivanoğlu[2], Yubin Huang[3], Ali Sheraz[2], Phillip S. Dobson[4], Jonathan M. R. Weaver[4], Pascal Gehring[3], T. Serkan Kasırga[1,2*]

[1] Bilkent University UNAM – Institute of Materials Science and Nanotehcnology, Bilkent 06800 Ankara, Turkey

[2] Department of Physics, Bilkent University, Bilkent 06800 Ankara, Turkey

[3] IMCN/NAPS, Université Catholique de Louvain (UCLouvain), 1348 Louvain-la-Neuve, Belgium

[4] James Watt School of Engineering, University of Glasgow, Glasgow G12 8LT, U.K

*Corresponding Author: kasirga@unam.bilkent.edu.tr

1. **Theoretical prediction of the substrate-effect induced Seebeck coefficient difference in MoS₂**

As discussed in the main text, we assume that the Seebeck coefficient S in thermal equilibrium is composed of contributions from the energy-dependent diffusion ($S_N$), scattering ($S_\tau$) and the phonon-drag ($S_{pd}$), so that $S = S_N + S_\tau + S_{pd}$. Here, $S_N$ and $S_\tau$ terms can be written from the Mott relation assuming that MoS₂ is in the highly conductive state and electrons are the majority carriers:

$$S_\tau = -\frac{\pi^2 k_B^2 T}{3e}\frac{\partial \ln\tau}{\partial E}\bigg|_{E=E_F} \text{ and } S_N = \pm\frac{k_B}{e}\left[\frac{E_F - E_C}{k_B T} - \frac{(r+2)F_{r+1}(\eta)}{(r+1)F_r(\eta)}\right]$$

As mentioned in the main text, $S_\tau$ is zero as $\tau$ is energy independent in the 2d limit. $S_N$ term is composed of constants related to material properties, scattering parameter $r$ and the Fermi integral of the $r$-th order: $F_r = \int_0^\infty \left[\frac{x^m}{e^{x-\eta}} + 1\right]dx$. The scattering parameters of 2d materials are listed in **Table 1**.[1,2] Here, as discussed in detail in Ref. [1], $r = 0$ adequately accounts for the acoustic phonon scattering and small deviations of experimental data from the calculated values is due to the other scattering mechanisms. As a result, at the room temperature $S_N$ for suspended MoS₂ ($10^{12}$ cm⁻²) is about -400 µV/K and for SiO₂ supported MoS₂ ($10^{13}$ cm⁻²) is about -200 µV/K.

**Table 1. Scattering parameters $r$ of 2d materials.**

| Scattering mechanism | $r$ |
|---|---|
| Charged Impurity Scattering | 3/2 |
| Acoustic Phonon Scattering | 0 |
| Intervalley Scattering | 0 |
| Strongly Screened Coulomb Scattering | -1/2 |

$S_{pd}$ term can be estimated from the theory of phonon-drag in semiconductors in the first order as $S_{pd} = -\frac{\beta v_p l_p}{\mu T}$ where, $v_p$ and $l_p$ are the group velocity and the mean free path of a phonon, $\beta$ is a parameter to modify the electron-phonon interaction strength and ranges from 0 to 1, and $\mu$ is the electron mobility, respectively. As the dominant charge carriers are electrons, $S_{pd}$ term has a negative sign. We use the parameters given in **Table 2**. Based on the values given in the table we obtain $S_{pd}^{SiO_2} = -230$ µV/K and $S_{pd}^{Sus} = -100$ µV/K.

The total $S = S_N + S_{pd}$ for suspended and SiO$_2$ supported parts can be calculated by adding both contributions. $S^{Sus} = -500$ µV/K and $S^{SiO_2} = -430$ µV/K. Of course, we consider this to be a rough estimate as we ignore charged impurity scattering and strongly screeded Coulomb scattering. Also there are certain errors associated with the measurement of the parameters used for the calculation of the Seebeck coefficients. However, overall, this calculation shows that the substrate induced effect must be present under right experimental conditions.

**Table 2. Parameters used for $S_{pd}$ calculation**

| Parameter | On SiO$_2$ (Ref. [3]) | Suspended (Ref. [3]) |
|---|---|---|
| $v_p$ | 7 10$^5$ cm/s | 7 10$^5$ cm/s |
| $l_p$ | 5 nm | 20 nm |
| $\mu$ | 5 cm$^2$/V.s | 50 cm$^2$/V.s |

2. **SPCM map on a gold electrode substrate engineered MoS$_2$ device**

Throughout the study we used indium contacted devices thanks to their rapid fabrication. To compare our indium device results, we fabricated gold contacted devices. **Figure S1** shows the optical microscope images and corresponding SPCM reflection and photocurrent maps. There is no qualitative difference between the indium contacted devices and gold contacted device in the substrate-engineered photocurrent. Despite IV measurement is collected from 0.25 to -0.25 V its rectifying behaviour can be observed. Power dependence of the photocurrent from the substrate engineered junction is also comparable to the one reported in indium contacted devices.

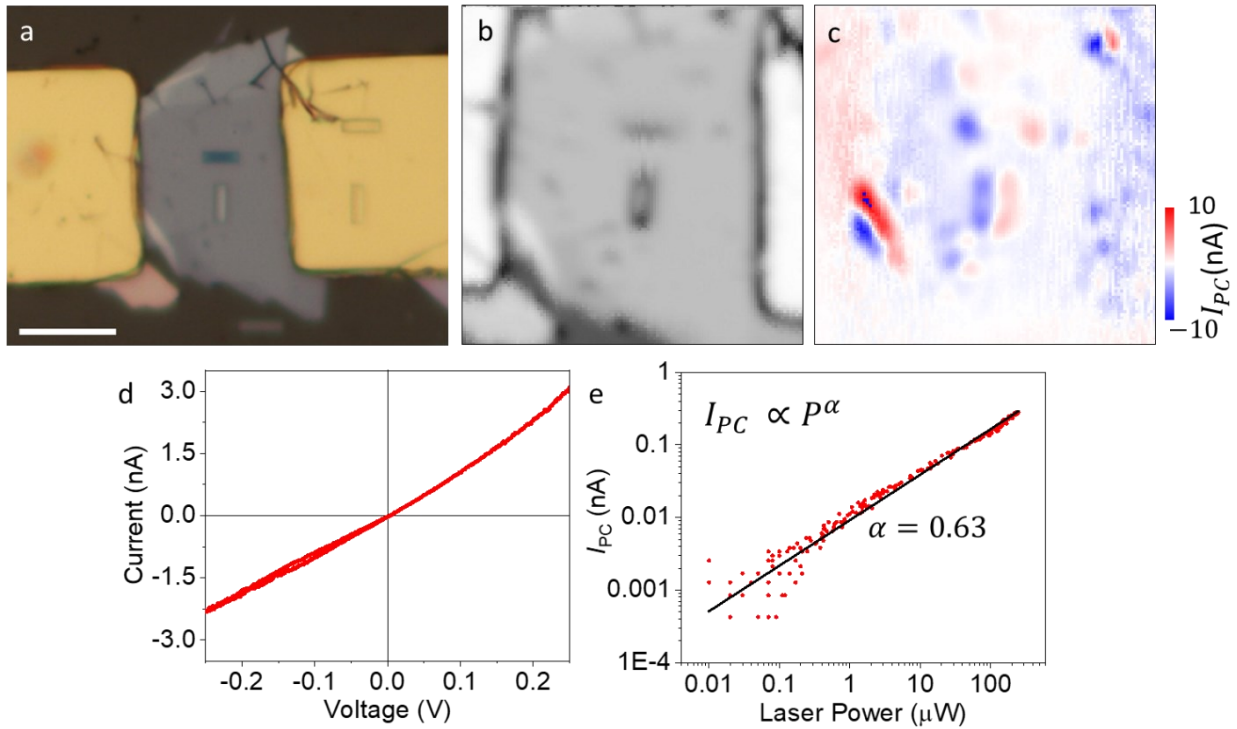

**Figure S1 a.** Optical microscope micrograph of a gold contacted substrate engineered MoS$_2$ device is shown. Scale bar is 10 µm. **b**. SPCM reflection map and **c**. photocurrent map. **d**. IV curve shows signs of rectifying nature of the contacts. **e**. Power dependence of the photocurrent at one of the side of the junction is plotted in a log-log graph and the exponent is about 0.63.

## 3. Scanning photovoltage microscopy, AFM and KPFM measurements on a parallel trench device

**Figure S2** shows an MoS$_2$ device fabricated on trenches drilled on sapphire with different depths. We performed SPVM, AFM and KPFM Measurements. First, AFM measurements show that the crystal is stuck to the bottom of the 100 nm deep trench (**Figure S2b**). For the rest of the trenches the flake bulges about 10 nm above the surface (**Figure S2c**). AFM height trace map also reveals a peculiar wrinkle formation over the suspended part of the flake.

In this measurement we operated the scanning microscope at photovoltage mode. **Figure S2d** shows the reflection map and the corresponding photovoltage map. The bipolar response is evident with slightly lower positive signal in some of the trenches. This asymmetry can be explained by lower heating of one side of the samples due to the scan direction. One important observation that agrees well with the photothermoelectric photoresponse is that the 100 nm trench shows very small photovoltage as compared to other trenches.

**Figure S2e** shows the KPFM and AFM profiles. The suspended part of the crystal has 60 meV lower surface potential difference. This is consistent with other KPFM measurements. The lower panel shows the variation in the height over the wrinkles. The workfunction is calculated with a calibrated tip and it follows the wrinkles of the sample. However, the difference in the SPD is not due to the variations in the height profile of the crystal. The change in the workfunciton is a good indication of the changes in the electronic landscape of the device upon suspension. Small variations along the wrinkles are also expected due to formation of varying stress regions along the crystal.

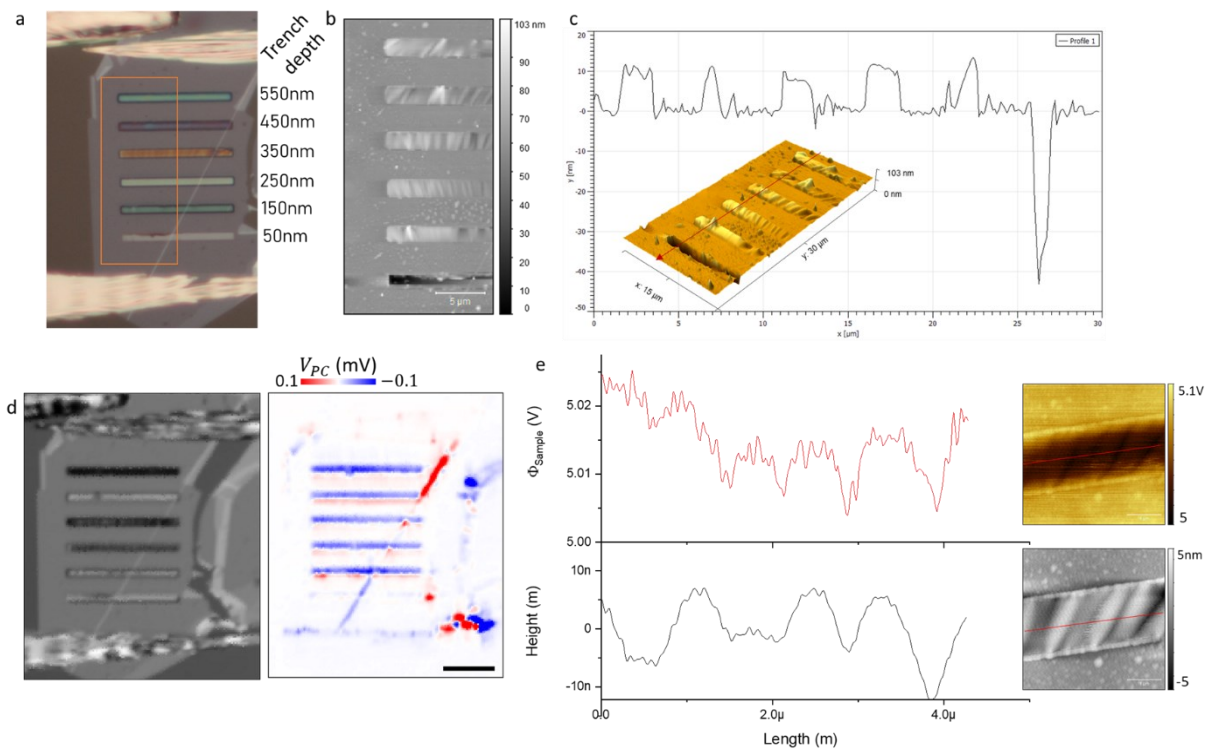

**Figure S2 a.** Optical microscope image of the device with trench depths labelled next to it. **b**. AFM height trace map and **c**. line trace taken along the height trace. The bulge of the crystal over the trenches is clear. **d**. Reflection and photovoltage maps obtained by operating the scanning microscope in photovoltage mode. Scale bar is 5 μm. **e.** Left panel shows the workfunction and height taken over the red lines marked on the maps given in the right panel. The variation of the workfunction along the trench is very small and correlated with the wrinkles of the crystal.

4. **SPCM maps taken at different laser wavelengths and incidence polarization**

We used three different wavelengths, 406, 532 and 633 nm, in our experiments all of them which are at an energy larger than the band gap of $MoS_2$. **Figure S3** shows the SPCM results collected with different laser wavelengths. Also, polarization dependence of the photocurrent measured at each end of the trench as well as a point over the contact is given in **Figure S3f**. There is no polarization dependence of the photocurrent. This shows that The effect is not due to built in polarization fields.

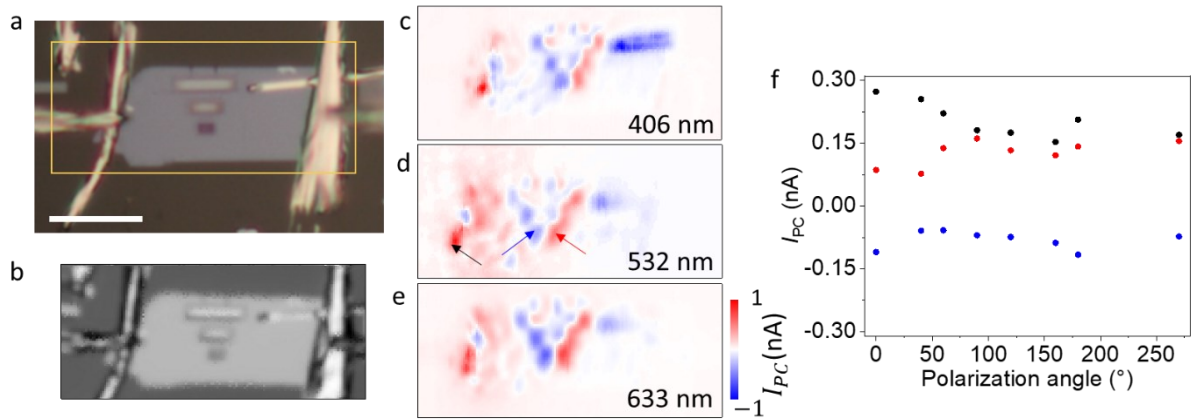

**Figure S3 a.** Optical microscope image of a two-terminal substrate-engineered $MoS_2$ device with different trench widths. Scale bar is 10 μm. **b.** SPCM reflection map of the region marked with yellow rectangle in **a**. **c**, **d** and **e** show photocurrent map taken at different wavelengths. At each run laser power is set to ~40 μW. The measured signal in all three measurements are very close and the overall photocurrent features are the same. **f.** Incident polarization of the 633 nm laser is rotated and $I_{PC}$ is measured at three different points marked by colored arrows on **d**, black dots- near contact, red dots- at the positive side and blue dots- at the negative side of the trench. There is no polarization dependence of the measured photocurrent at the three points where photocurrent is measured.

5. **Finite element simulation of a substrate modified thermoelectric junction**

To understand how a substrate modified thermoelectric junction would behave depending on how the contacts are configured, we performed finite element analysis simulations using COMSOL Multiphysics. An irregularly shaped crystal is modelled over a substrate with a hole and voltage at a floating terminal is measured with respect to different laser positions. The observed pattern agrees with our measurements. **Figure S4** shows the thermoelectric emf generated and temperature distribution maps.

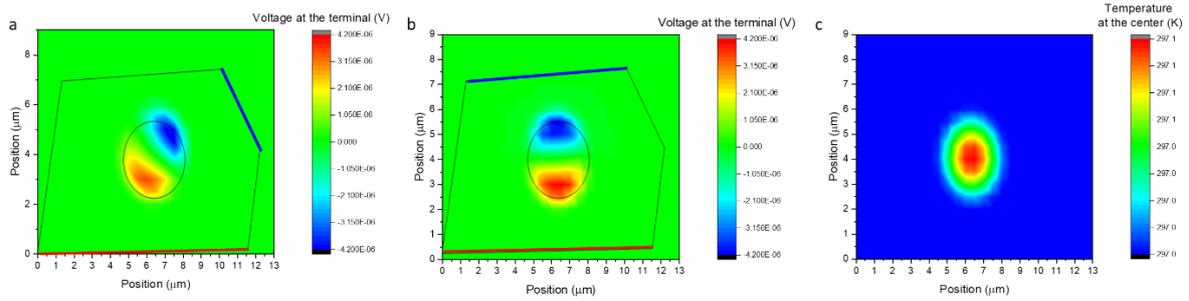

**Figure S4 a.** An arbitrary crystal is modelled over a SiO$_2$ substrate with a hole. The outline drawn over the voltage distribution map shows the outline of the crystal and the outline of the hole. The red line indicates the ground terminal and the blue line indicates the floating voltage terminal where the photothermoelectric emf is measured from like in the experiments. **b.** For comparison, another terminal is simulated as the floating terminal. **c.** Temperature distribution vs. laser position is shown. As clear, the maximum temperature rise is achieved at the center of the hole.

### 6. KPFM on h-BN supported and suspended MoS$_2$

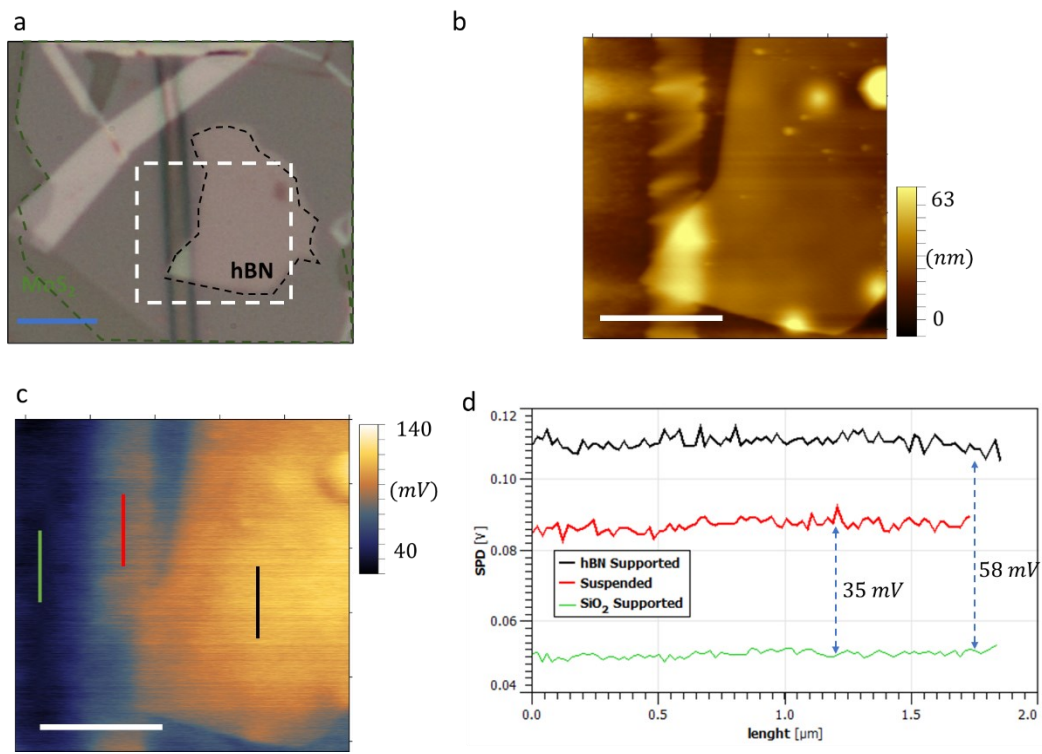

**Figure S5 a.** Optical micrograph of an MoS$_2$ crystal partially suspended over a trench and partially supported on h-BN (outlined by blacked dashed lines). White dashed square shows the AFM region. **b.** AFM height trace map and **c.** corresponding SPD map is given. **d.** SPD line traces from the colored lines in **c** are plotted. The difference between the SPD in h-BN supported, suspended and SiO$_2$ supported parts are evident. Scale bars are 2 µm.

### 7. Gate dependent measurements

We performed gate dependent SPCM measurements both on suspended and h-BN supported MoS$_2$ devices. In both cases, we used 1 µm SiO$_2$ coated Si wafers. Si is used as the back gate in both device configurations. We reported the h-BN supported junctions in the main text as devices over holes showed significant change upon application of negative gate bias. **Figure S5** shows the degradation of the suspended device. After application of a few volts the device irreversibly shows a contrast change starting from the edges of the hole. We fabricated a long trench with open ends to see if the trapped air within the hole is causing the observed contrast change. However, same contrast change is observed after applying negative gate voltages. We observe that the contrast change starts from near the hole and expands from there. At the moment we are not fully aware of the reasons leading this contrast change. We consider that the release of the adsorbed molecules on the surface of the substrate under large negative gate voltages lead to such degradation.

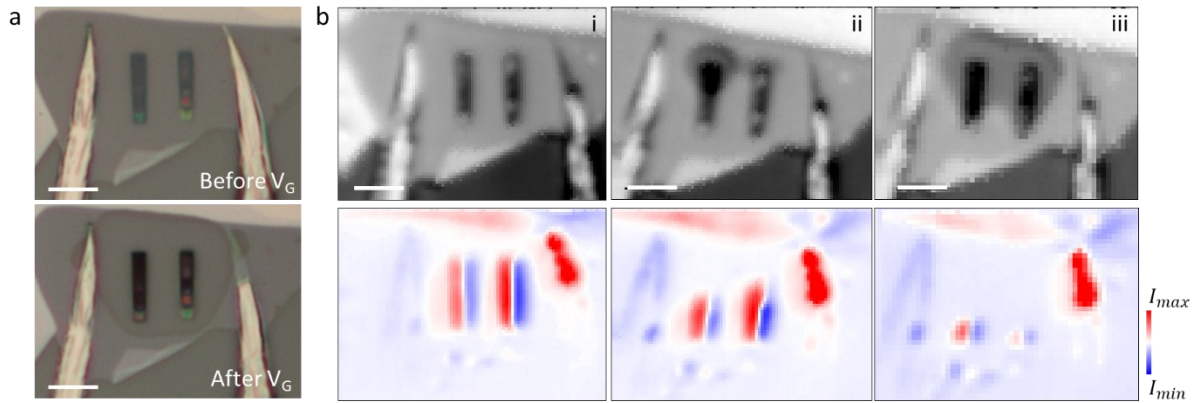

**Figure S6 a.** Optical microscope micrograph of indium contacted MoS$_2$ on SiO$_2$/Si with stair-like holes before and after application of gate voltages down to $V_G = -20\ V$. Lower panel shows a clear contrast change around the holes extending to the indium contacts. **b.** SPCM maps taken at $V_G = 0\ V$ with 532 nm of 86 µW on sample: (i) before gating, (ii) after $V_G = -15\ V$ scan and (iii) after the scan in (iii). $I_{max} = 6.5$ nA and $I_{min} = -6.5$ nA. Scan starts from top left corner to the bottom left corner with progressing to the right in raster scan pattern. Scale bars are 5 µm.

To prevent the sample degradation problem under large negative gate voltages, we coated the substrate surface with 5 nm thick Al$_2$O$_3$ using atomic layer deposition (ALD) method after milling the holes with FIB. Then, the device is fabricated over the ALD coated surface. The device didn't show any sign of degradation and produced pronounced photoresponse. Measurements from the device is given in **Figure S6**. Although the device exhibits the expected gate dependent response, as discussed in the main text, there is no carrier inversion induced reduction in the photovoltage due to the Fermi level pinning.

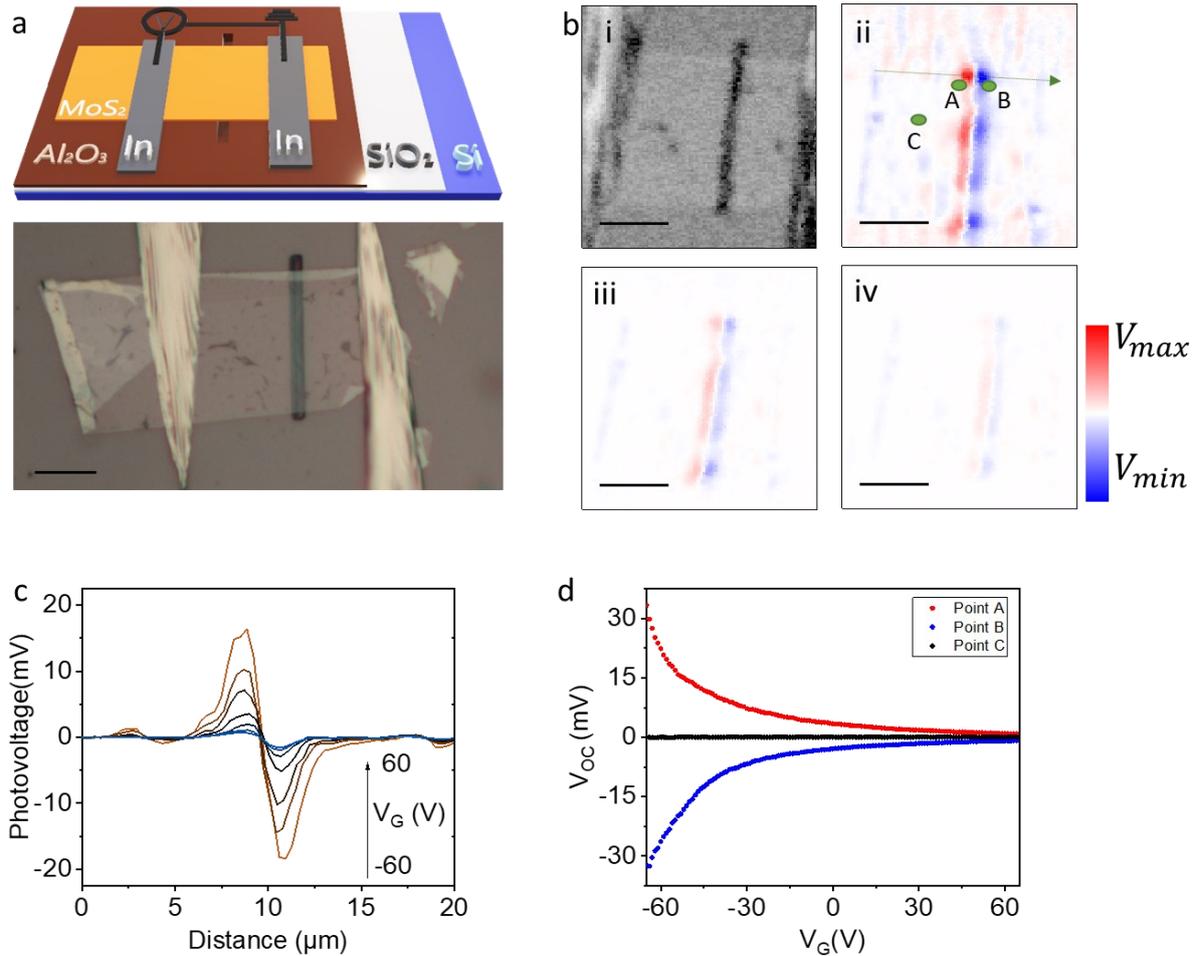

**Figure S7 a.** Schematic of the device along with the optical image is shown. The sample is coated with 30 nm thick $Al_2O_3$ to passivate the $SiO_2$ surface and to minimize the pinholes. Scale bar is 10 μm. **b.** Photovoltage map collected in DC mode without the Lock-in amplifier and chopper. **i** is the reflection map, and photovoltage maps at **ii** is the $V_G$ =-60 V, **iii** $V_G$ = 0 V, **iv** $V_G$ =60 V. Here, $V_{max}$ = 20 mV and $V_{min}$ = -20 mV. **c.** Photovoltage line trace taken along the dashed arrow given in **b-ii**. Large signal corresponds to the more negative gate voltages. **d.** Photovoltage data collected from points indicated on **b-ii**. This sample showed no Seebeck coefficient inversion due to possible Fermi level pinning induced by the substrate as discussed in the main text.

H-BN supported devices performed better and showed no sign of such a contrast change. **Figure S7** shows the reflection and the photocurrent maps reported in the main text and the photocurrent from point 2 and 3 subtracted from point 4, marked on the photocurrent map. Both junctions of the h-BN show almost identical response under gate voltage (point 3 data is multiplied by -1 for viewing convenience).

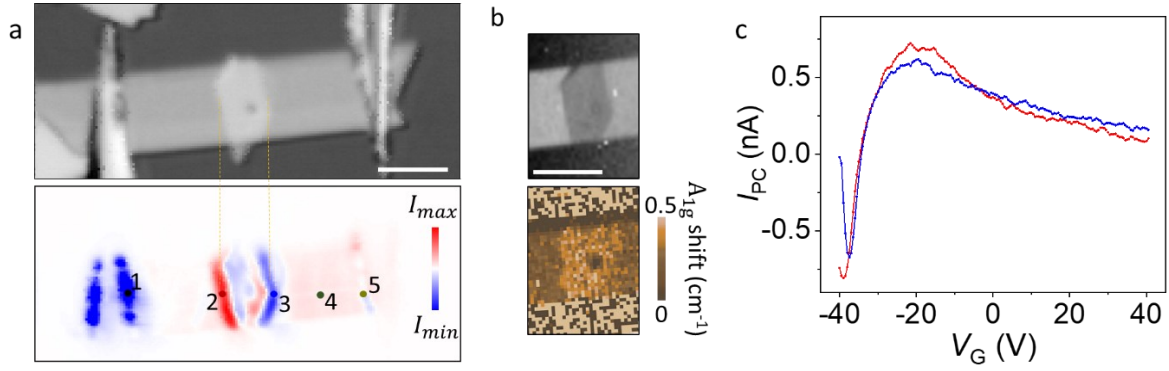

**Figure S8 a.** Same figure from the main text is copied here for convenience. **b.** Raman intensity map and the $A_{1g}$ peak shift map is given. **c.** Gate dependent signal from point 4 is subtracted from the gate dependent data from point 2 (red curve) and point 3 (blue curve). Blue curve is multiplied by -1 for viewing convenience.

## 8. Scanning Thermal Microscope Calibration and Seebeck variation estimation

The Scanning Thermal Microscope (SThM) measurements were performed on a commercial Bruker Icon instrument with a VITA-GLA-DM-1 probe. The probe, consisting of the silicon nitride lever with a Pd heater/thermometer has been calibrated on a hot plate to relate the temperature to its electrical resistance. The calibration curves are shown on figure S9.

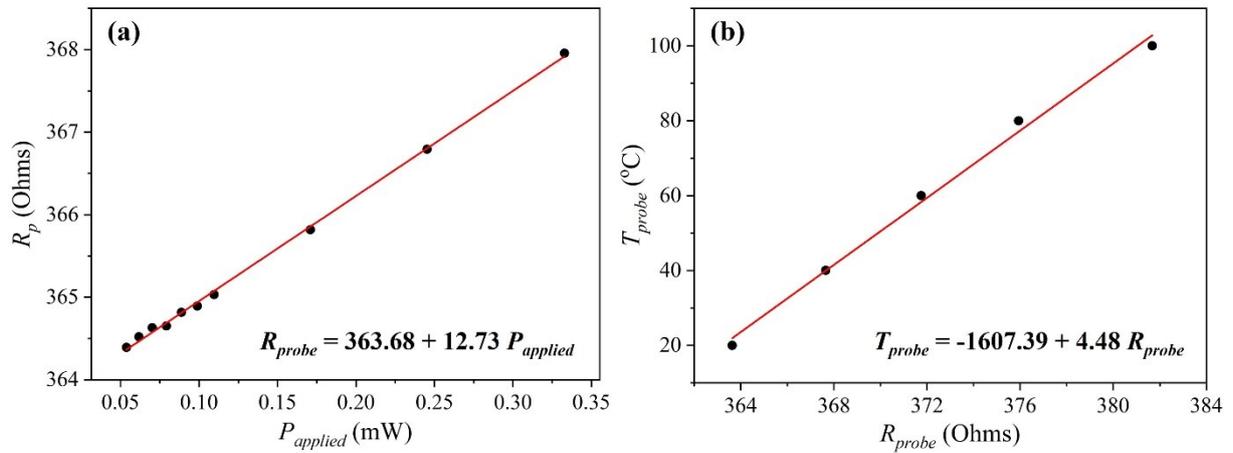

**Figure S9 a.** SThM probe calibration of the electrical resistance with the supplied power. **b.** Temperature as a function of electrical resistance

As described elsewhere[4,5], the probe is part of a modified Wheatstone bridge which is balanced at low voltage. During the measurements, we applied a combined AC (91 kHz) and DC bias on the bridge which heats the probe and creates a bridge offset that directly measures the probe heater temperature. For most experiments, we applied 1mW on the probe creating a $\Delta T$ of 50 ± 2 K, when the probe was far away from the sample.

When the SThM tip is brought into contact with the devices, it locally heats the materials below its apex. While the probe scans the surface, the device open circuit voltage is recorded and amplified via a SR830 voltage preamplifier. This voltage is referred to as the thermovoltage. We excluded any shortcut between the probe and the device as no leakage current could be measured between the probe and both contacts.

The thermovoltage can be written analytically as[6,7],

$$V_{th}(x) = -\int_A^B S(x)\frac{\partial T}{\partial x}(x)dx$$

where $S(x)$ is the position dependent Seebeck coefficient and $\frac{\partial T}{\partial x}(x)$ is the position dependent temperature gradient. Both are integrated over the whole device length from A to B.

As shown elsewhere[6,7], it is possible to deconvolute the Seebeck coefficient from the temperature gradient. This however requires a precise estimation of the temperature gradient and thus the sample temperature rise under the tip, $\Delta T_{sample}$.

As we know the probe temperature far away from the sample (50 ± 2 K) and we monitor its temperature via the Wheatstone bridge, we know that the probe temperature in contact with the sample is 43.8 ± 4 K. The probe cooling occurs because of several heat transfer mechanisms[4,5] (solid-solid conduction, air conduction, water meniscus, …).

For those probes, the Pd heater is however distributed over the whole triangular shaped silicon nitride tip[4,5]. This implies that the tip temperature and probe temperature are different. We turned to finite element modelling (COMSOL Multiphysics) to estimate the tip temperature over the $MoS_2$ suspended and supported sample. Figure S10 shows the overall simulated probe and sample.

We used reported values for the in-plane and out-of-plane $MoS_2$ thermal conductivity as well as for the $MoS_2$-glass interface conductance. Reported values vary greatly in literature[8–16]. However, to the best of our knowledge, for a thick sample (>10 layers), the values are on the order of 30 $Wm^{-1}K^{-1}$ for the supported in-plane, 60 $Wm^{-1}K^{-1}$ for the suspended in-plane and 3 $Wm^{-1}K^{-1}$ for the cross-plane conductivities. For the substrate interface conductance, we used 1 $MWm^{-2}K^{-1}$.

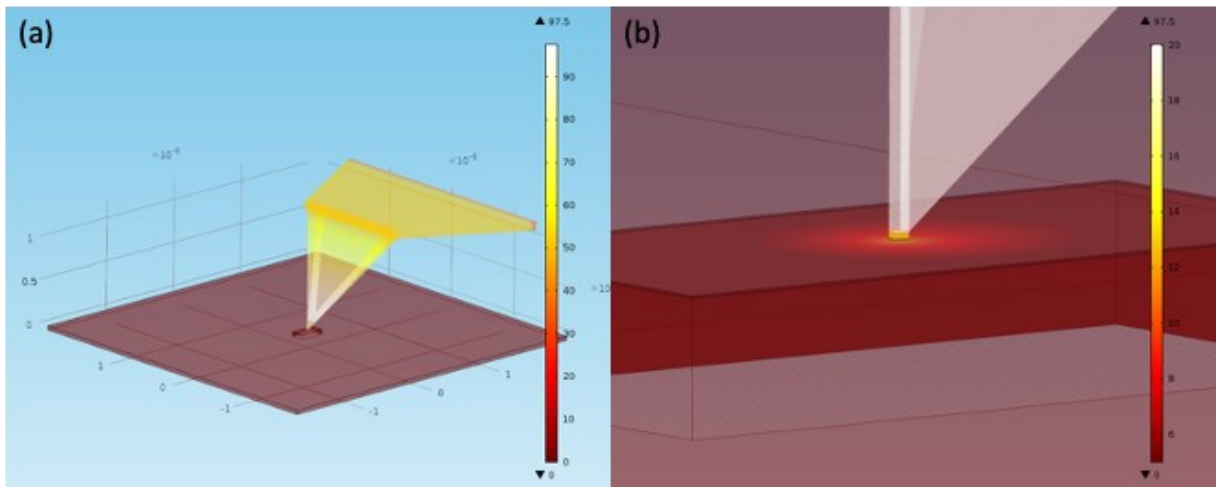

**Figure S10 a.** Finite element model for the SThM probe on a $MoS_2$ suspended sample. **b.** Zoomed-in view of the model where the temperature gradient is visible on the sample surface.

Using those material parameters, we estimated a ratio between the probe temperature and the tip apex temperature of 4.9. The model also accounts for the tip-sample thermal resistance. This method and model were experimentally confirmed elsewhere[4,5,17]. Taking these into consideration, we obtain a sample temperature rise $\Delta T_{sample}$ of 7.4 ± 0.7 K. This gives a Seebeck variation of 72±10 µVK$^{-1}$.